\documentclass[final,3p,times,numbers]{elsarticle}
\usepackage[english]{babel}
\usepackage{amssymb}
\usepackage{amsmath}
\usepackage{xcolor}
\usepackage{booktabs}
\usepackage{url}
\usepackage{hyperref}
\usepackage{silence}
\usepackage{lmodern}
\usepackage{url}
\usepackage{hyphenat}
\usepackage{lmodern}
\usepackage{xcolor}
\usepackage{float}
\usepackage{caption}
\usepackage{graphicx}
\usepackage{amsmath}
\usepackage{epsfig}
\usepackage{placeins}
\usepackage{booktabs, xcolor}
\journal{Blockchain: Research and Applications}
\begin{document}
\begin{frontmatter}
%% Keywords
\begin{keyword}
Bitcoin User Network \sep Blockchain Network Analysis \sep Complex Networks \sep PageRank \sep HITS
\end{keyword}




\title{Concentration Within Distribution: Unmasking Bitcoin's Structural Centralization Through Network Science} %% Article title


\author[label1,label2]{Myriam Nonaka \corref{cor1}}%% Author name
\cortext[cor1]{Corresponding author:}
\ead{myriam.nonaka@urjc.es, mnonaka@citedef.gob.ar}
\author[label1]{F. Javier Marín-Rodríguez}
\author[label1,label3]{Alexander Jiricny}
\author[label1,label5]{Miguel Romance} 
\author[label1,label5]{Regino Criado}
\author[label5,label7]{Sergio Iglesias-Pérez}
\author[label5,label6]{Alberto Partida} %% Author affiliation

\affiliation[label1]{organization={Department of Applied Mathematics, Materials Science and Engineering, and Electronic Technology, Laboratory of Mathematical Computation on Complex Networks and their Applications, Rey Juan Carlos University},%Department and Organization
            addressline={Calle Tulipán s/n}, 
            postcode={28933},
            city={Móstoles}, 
            state={Madrid},
            country={Spain}}
            
\affiliation[label2]{organization={Instituto de Investigaciones Científicas y Técnicas para la Defensa (CITEDEF-CONICET)},%Department and Organization
            addressline={San Juan Bautista de Lasalle 4397}, 
            postcode={1603},
            city={Villa Martelli}, 
            state={Buenos Aires},
            country={Argentina}}                       
\affiliation[label3]{organization={Universidad Tecnológica Nacional (UTN-FRBA)},%Department and Organization
            addressline={Av. Medrano 951}, 
            postcode={1179}, 
            city={CABA},
            country={Argentina}}
            
\affiliation[label5]{organization={Data, Complex Networks and Cybersecurity Sciences Technological Center,
Rey Juan Carlos University},%Department and Organization
            addressline={Plaza de Manuel Becerra 14},
            postcode={28028},
            city={Madrid}, 
            country={Spain}}

\affiliation[label7]{
            organization={Department of Computer Science and Technology, Universidad Internacional de La Rioja},
            addressline={Avenida de la Paz 137},
            postcode={26006},
            city={Logroño},
            state={La Rioja},
            country={Spain}
}
            
\affiliation[label6]{organization={Department of Computing and Technology, School of Architecture, Engineering, Science and Computing (STEAM), Universidad Europea de Madrid}, 
            addressline={Campus de Villaviciosa, Calle Tajo s/n}, 
            postcode={28670},
            city={Villaviciosa de Odón},
            state={Madrid},
            country={Spain}}

%% Abstract
\begin{abstract}
We construct the Bitcoin User Network (BUN) directly from raw blockchain data up to late 2025, which allows us to explore its mesoscopic properties and trace its temporal evolution. In particular, we analyze the structure of connected components and directed assortativity through the four variants of Newman’s coefficient, implemented via custom algorithms and a dedicated database. Building on this, to characterize the distribution of structural influence, we introduce direction-sensitive centrality measures based on PageRank and HITS, which provide a complementary global analysis of the BUN and reveal a persistently unequal and increasingly core–periphery structure. In addition, we complement the structural analysis with a study of Bitcoin’s price volatility using high-frequency market data. Overall, our results reveal a clear pattern of concentration within distribution: although the protocol is decentralized by design, the emergent user network evolves toward an asymmetric mesoscopic structure that indicates the existence of a few large-scale connected components that function as the critical backbone of the system.\\

\end{abstract}
\end{frontmatter}    
\section{Introduction}
\label{sec1}
%% Labels are used to cross-reference an item using \ref command.
The original paper authored by Satoshi Nakamoto \cite{Nakamoto2008} introduces the concept of Bitcoin, the world’s first cryptocurrency. Emerging during the 2008 global financial crisis, Bitcoin is designed as a radical departure from traditional financial structures. The fundamental premise behind Bitcoin is the decentralization of money, where no single authority controls the money supply. It ensures transparency through a publicly verifiable ledger and promotes trustlessness by removing the need for intermediaries, allowing transactions without going through financial institutions. In this work, we adopt the conventional distinction where “Bitcoin” (capitalized) refers to the network, protocol, and overall system architecture, while “bitcoin” (lowercase) or BTC denotes the currency unit transacted a single network.\\

In an abstract sense, a transaction is a set of input and output addresses. To implement a secure decentralized network that functions without the need for a trusted third-party entity, Bitcoin utilizes a transaction history based on cryptographic proof and generates a "chain" of blocks (blockchain), whose "concatenation" increases computational difficulty as time progresses and reduces the probability of a cyberattack. 
Unlike traditional financial institutions, Bitcoin enables anyone to publicly inspect all transactions while ostensibly preserving user anonymity. However, this anonymity is, in fact, pseudonymity, as direct user identification, although difficult, is not impossible \cite{REID}. The transparency of the network, where all transactions are publicly recorded, creates opportunities for
analysis. Researchers have developed heuristics \cite{Harrigan, Meiklejohn} to identify
common ownership patterns and link addresses to the same users by analyzing
transaction structures, timing patterns, and behavioral signatures in how Bitcoin manages token exchanges.\\

Early comprehensive analyses \cite{Bohme,Partida}, framed Bitcoin as a socio-technical system that combined the features of the currency, commodity, and payment infrastructure, while also highlighting the challenges related to scalability, governance, and regulation. However, more recent work increasingly situates Bitcoin in a single domain of financial economics. As summarized in a recent survey \cite{Kang}, Bitcoin is widely studied as a financial asset characterized by relative hypervolatility (see Section \ref{sec:hyper}), asymmetric dependencies, and weak correlations with traditional asset classes. This shift reflects a growing consensus in the literature that Bitcoin operates less as a pure medium of exchange and more as a digital reserve asset, as well as a speculative and potentially hedging instrument within global financial markets.  As time progresses, Bitcoin evolves quickly and it is therefore  necessary to work with updated data. Its transactions grow to be more voluminous and complex, and user behavior on the network becomes increasingly relevant for its study to better understand the cryptocurrency market. Is Bitcoin's volatility driven by external factors, or is it inherent to the system?\\

\subsection{Summary of Contributions}
In this work, we construct the Bitcoin User Network (BUN) from 2011 up to late 2025 using a complete full-node dataset and custom address-clustering heuristics aligned with the methodology of Vallarano et al \cite{Vallarano}. This enables us to replicate and extend canonical mesoscopic results on connected components and directed assortativity, providing a unified temporal characterization of the BUN over an almost 15-year period. Beyond this extension, we incorporate direction-sensitive centrality measures—PageRank and HITS—to quantify the concentration of incoming and outgoing structural influence, revealing persistent inequality and pronounced core–periphery patterns a single network. Our analysis also uncovers a phase of transient mesoscopic turbulence in the strongly connected core during 2023–2024, interrupting an otherwise long-term regime of topological stationarity. Finally, we complement the structural study by examining Bitcoin’s hypervolatility using high-frequency spot price data from 2020 up to late 2025, providing an orthogonal perspective on market dynamics relative to the underlying network evolution.

\subsection{Terminology and Structural Definitions}

To ensure clarity and self-containment, we briefly define the structural concepts used throughout this work. The term \textit{mesoscopic structure} refers to properties of the network that emerge at an intermediate scale between microscopic (address-level) behavior and macroscopic (global) statistics. These include the size and dynamics of large connected components, assortativity patterns, and the distribution of centrality values across users. \\

We distinguish between weakly and strongly connected components. A \textit{weakly connected component} (WCC) is a maximal set of nodes that remain connected when edge directions are ignored, whereas a \textit{strongly connected component} (SCC) requires mutual reachability through directed paths (edges between the nodes). The largest weakly connected component (LWCC) typically contains most of the users in the network, while the largest strongly connected component (LSCC) represents the mutually reachable core of the directed graph. \\

When describing hierarchical organization, we use the notion of a \textit{core–periphery structure}, in which a dense and highly influential subset of nodes interacts asymmetrically with a large, sparsely connected periphery. Throughout this work, we refer to the LWCC as the \textit{backbone} of the network when it carries the vast majority of users and transaction paths. \\

We also distinguish between three forms of stability. \textit{Topological stability} refers to the stability of the number of connected components; \textit{mesoscopic stability} refers to the stability of the relative size of major components; and \textit{dynamical stability} refers to the temporal smoothness of structural indicators such as assortativity or centrality. Conversely, dynamical instability denotes short-term fluctuations caused by bursts of transactional activity. \\

The term \textit{assortativity} is used to describe the correlation between the degrees of nodes connected by a directed edge. Positive assortativity indicates that nodes tend to connect with others of similar degree, while negative assortativity (\textit{disassortativity}) indicates that high-degree nodes preferentially connect with low-degree nodes. In directed networks, we consider four variants based on combinations of in-degree and out-degree. Persistent disassortativity reflects a hierarchical flow pattern in which central hubs interact asymmetrically with peripheral users. \\

Finally, for each user, we define the \textit{in-degree} as the number of incoming transactions and the \textit{out-degree} as the number of outgoing transactions. Structural inequality is quantified using the Gini coefficient, applied here to centrality distributions such as PageRank and HITS.

\section{Data processing and considerations for the assembly of the BUN}\label{sec:data}
Bitcoin, unlike euros or dollars, does not have a physical form. Rather, it is made up of bits of information. Some studies analyze how an intangible currency becomes tangible and why it has monetary value despite not having governmental authority to create and control it, mentioning the advantages of cryptocurrencies in terms of security and general acceptance by the public \cite{Yermack}.\\

The advantage of Bitcoin is that the network allows access to the entire transaction history and validates the blockchain through Bitcoin Core, which is open-source software \cite{BitcoinOrgBitcoinCore} and serves as an official client that connects to the network, allowing users to receive, send, and manage transactions. This feature of the Bitcoin network guarantees the availability of verified data and allows direct queries about transactions within blocks, without relying on third parties. This information implies a large volume of data, currently exceeding hundreds of gigabytes (GB). As of November 2025, the dataset size corresponding to a full node on the Bitcoin network is estimated to be approximately 700 GB, reflecting the cumulative expansion of the blockchain since its early stages of development.\\

\begin{figure*}[!ht]
\centering
\captionsetup{justification=centering}
\includegraphics[width=0.8\textwidth]{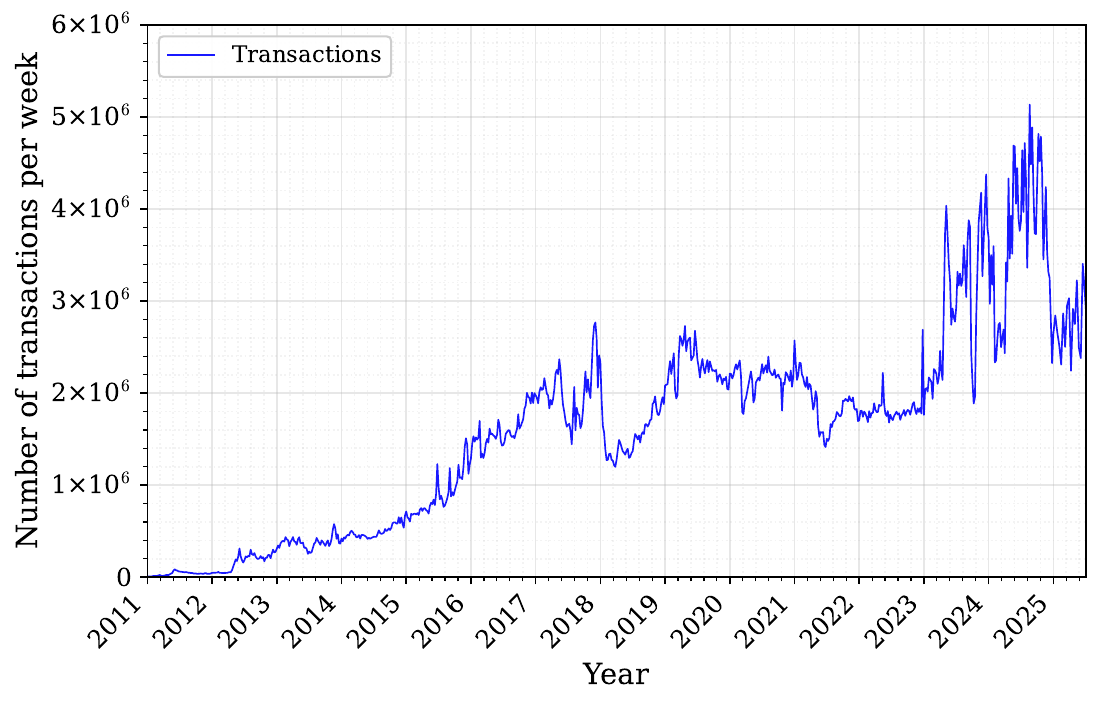}
    \caption{Weekly number of Bitcoin transactions from 2011 to 2025, showing long-term growth from under 1 million to peaks near 5 million transactions per week.}
    \label{fig1:BTC network evolution}
\end{figure*}

\subsection{Considerations of data processing} 

The processing and storage of Bitcoin
transaction data, due to the continuous flow of information, present technical challenges that must be considered. Figure \ref{fig1:BTC network evolution} shows the volume of transactions and the size of the Bitcoin data from 2011 to November 2025, based on our data. As illustrated, the number of transactions does not exceed 1 million per week between 2011 and 2014. However, from 2015 to 2017, this figure nearly triples. Between 2017 and 2025, the data show fluctuations, with a maximum peak exceeding 5 million transactions per week. These data reflect the increasing transaction volume and the higher amount of data handled by the Bitcoin network over time. Obtaining data from Bitcoin Core requires efficient reading of each block and transaction, extraction of inputs, outputs, and associated metadata.\\

Our analysis is based on the following resource requirements for data storage and processing:

\begin{itemize}
       \item Software: GNU Linux UBUNTU SERVER, Python 3.9.23, psql (PostgreSQL) 13.20, and Bitcoin Core RPC client version v27.0.0.
       \item Hardware: We use an HP Z840 server with Intel Xeon E5-2699 (44 cores), NVIDIA Quadro P5000 with 128 GB of RAM and 2TB SSD memory.
Disk usage was optimized through incremental parsing, selective indexing, and database normalization strategies, allowing all blockchain and derived datasets to be managed within same storage environment without pruning.
\end{itemize}  
\subsection{Bitcoin User Network (BUN) }

\begin{figure*}[ht]
\centering
\captionsetup{justification=centering}
\includegraphics[width=0.8\textwidth]{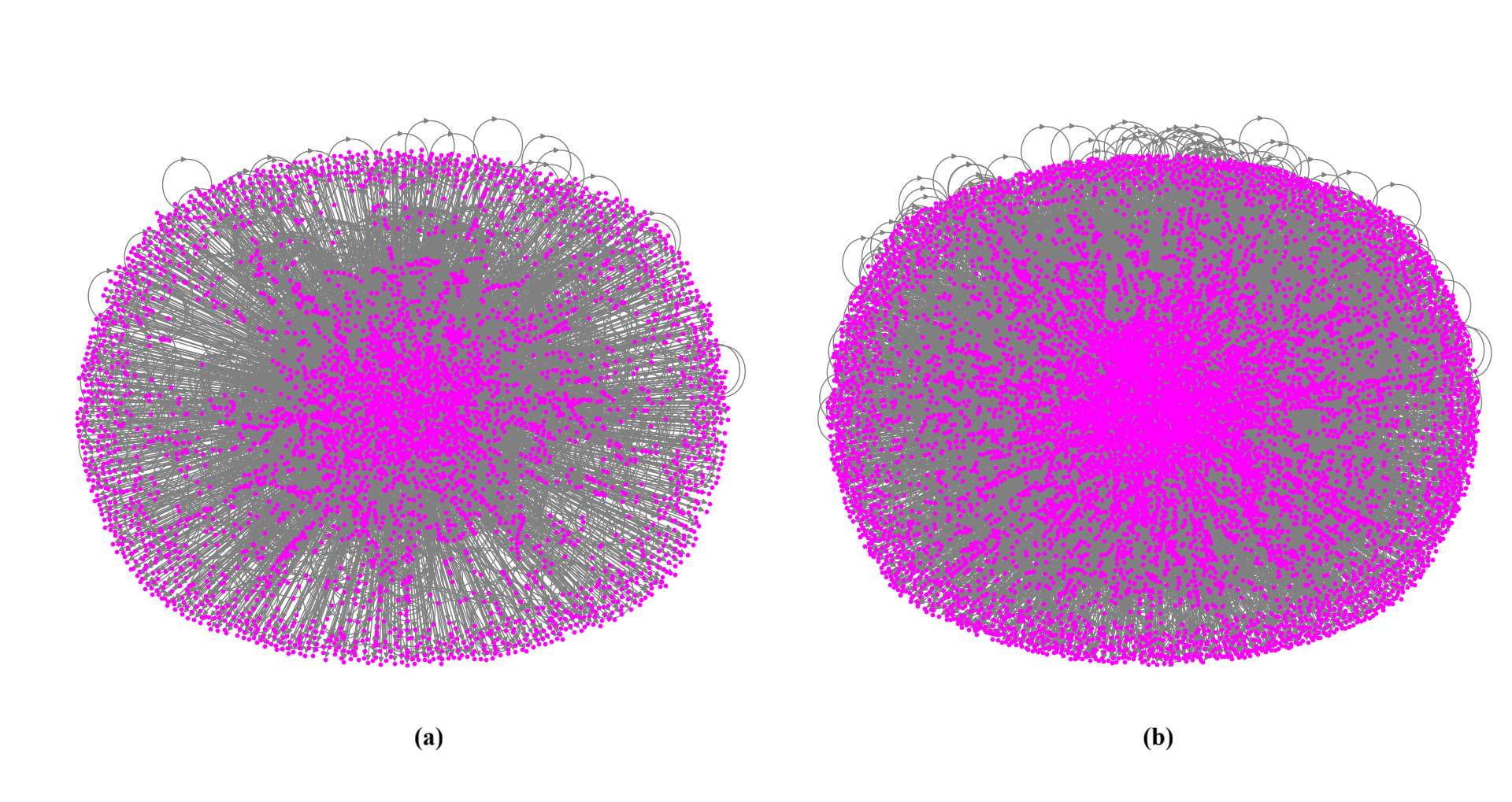}
    \caption{Snapshots of the BUN at different time intervals. (a) Early 2011 (week 104), showing initial network formation. (b) Ten weeks later (week 114), illustrating the rapid expansion of the network}
    \label{fig:2Snapshots}
\end{figure*}

Due to the privacy mechanisms of Bitcoin, we only have access to transactions and addresses, which are not linked to specific users. As a result, it is not possible to explicitly determine which addresses belong to the same user. However, we implement the heuristics \cite{Harrigan,Meiklejohn}previously proposed in the literature to approximately reconstruct groups of addresses controlled by the same user. This reconstruction is essential for analyzing user behavior, tracking fund flows, studying mesoscopic properties, and conducting privacy research on Bitcoin, as it provides insights beyond a mere list of addresses and transactions. Figure \ref{fig:2Snapshots} illustrates the rapid expansion of the network by showing snapshots of the BUN at week 104 (2011) and ten weeks later.

\subsection{Applied heuristics }
To be able to identify users, we rely on the BUN of Vallarano \textit{et al.} \cite{Vallarano}. To do this, we consider the following heuristics to reconstruct the relationships between addresses and users:
\begin{itemize}
       \item Multi-input heuristic: When multiple addresses appear as input ($\text{v}_{in}$) in the same transaction, it is assumed that they belong to the same user.
       \item Change heuristic: In transactions with multiple outputs ($\text{v}_{out}$), we approximate the change output using a simplified value-based heuristic, assuming that in standard multi-output transactions the smaller output often corresponds to change, while the rest are treated as recipients, appropriately assigning new or existing users. We recognize that this approach does not capture all behavioral patterns present in modern transaction structures (see Section \ref{sub:limitations}).
\end{itemize}

We emphasize the reassignment of $v_{out}$ and $v_{in}$ for the construction of the BUN: if an address previously used as $v_{out}$ later appears as $v_{in}$ alongside other addresses, the groups are automatically merged, ensuring consistency in user identification. These heuristics allow us to construct the Bitcoin User Network over long temporal scales, prioritizing structural robustness and scalability over fine-grained behavioral reconstruction.\\

In Table \ref{TableTransactions} , we present an example of a transaction table that lists the input and output addresses for each transaction. We represent the transactions as $T_i$, where $i$ indicates the transaction number; as an example, we show three transactions ($T_1$, $T_2$ and $T_3$). 
The $v_{in}$ columns represent the input addresses involved in each transaction, 
while the $\text{v}_{out}$ columns indicate the output addresses. 
The symbol \textit{A} in the $\text{v}_{in}$ column denotes an address associated with a certain amount of BTC; 
in this example, it corresponds to 5~BTC in transaction $T_1$.\\

\begin{figure}
\centering
\captionsetup{justification=centering}
\includegraphics[scale=0.45]{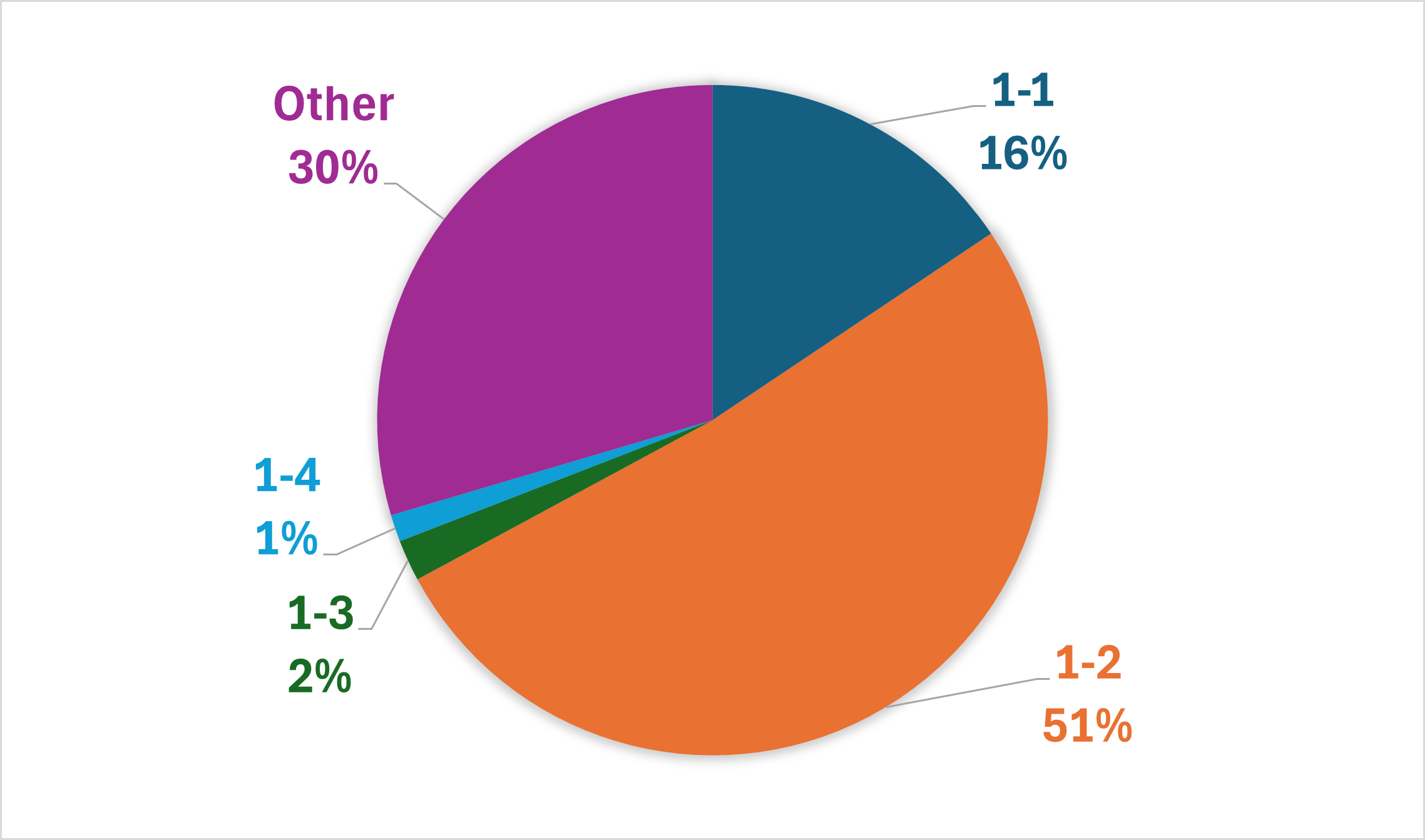}
    \caption{Distribution of Bitcoin transactions by number of inputs ($v_\text{in}$) and outputs ($v_\text{out}$) during week 700. 
``1-1'' indicates transactions with one input ($v_\text{in}=1$) and one output ($v_\text{out}=1$); 
``1-2'' indicates $v_\text{in}=1$ and $v_\text{out}=2$; 
``1-3'' indicates $v_\text{in}=1$ and $v_\text{out}=3$; 
and ``other'' represents all other combinations of $v_\text{in}$ and $v_\text{out}$.}
\label{fig:transacciones}
\end{figure}

\begin{figure}[ht]
\centering
\captionsetup{justification=centering}
\includegraphics[scale=0.80]{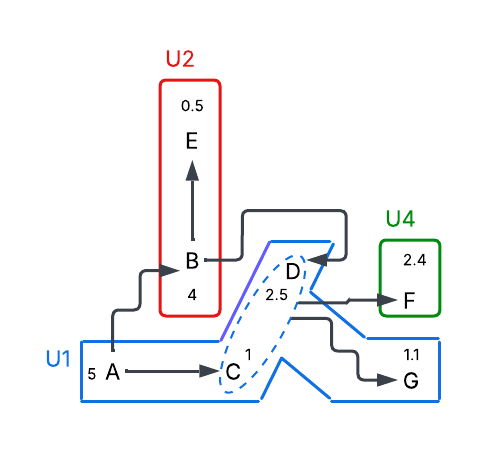}
    \caption{Representation of transactions and user groupings of Table \ref{TableAddress}. The arrows indicate transactions between addresses, while the colored blocks correspond to users grouped according to Table \ref{TableAddress}. User 2 (U2) is shown in red, user 4 (U4) in green, and user 1 (U1) in blue. The dashed line indicates the grouping of addresses that also belong to user 1. The numbers inside the blocks represent, as an example, arbitrary amounts of BTC transferred between transactions.}
\label{esquema}
\end{figure}

Based on this information, we show in Table \ref{TableAddress} how users are grouped, indicating which addresses are associated with each user, and Figure \ref{esquema} illustrates this grouping in a diagram. Note that user 3 (U3) does not have any assigned addresses, as it essentially corresponds to user 1 (U1). These heuristics, widely used in the literature, allow us to follow the methodological approach proposed by Vallarano \textit{et al.} \cite{Vallarano} for building the BUN, thereby ensuring coherence and comparability with their analytical framework.
 Figure \ref{fig:transacciones} shows, for week 700, the distribution of transactions by number of inputs ($v_{in}$) and outputs ($v_{out}$). Fifty-one percent of the transactions correspond to one input ($v_{in}$ =1) and two outputs ($v_{out}$ =2), forming the most frequent pattern and therefore making them particularly compatible with the heuristic we apply in this study. This pattern remains consistent throughout the weeks analyzed.\\

\begin{table}[ht]
\centering
\scalebox{0.8}{
\begin{tabular}{p{3cm}p{3cm}p{3cm}}%{l|}%{|>m{5cm}|>m{15cm}|}
%\rowcolor{orange}
\hline
 \textbf{Transaction number}  & \textbf{$v_{in}$ 
}& \textbf{$v_{out}$}\\\hline
T1&A=5&B=4,C=1  \\
\hline 
T2&B=3&D=2.5,E=0.5\\
\hline
T3&C=1,D=2.5&F=2.4,G=1.1\\
\hline
%\hline
\end{tabular}}
\caption{Example of three transactions (T1,T2 and T3) showing input ($v_{in}$) and output ($v_{out}$) addresses. The A in vin represents an input address with a BTC value (5 BTC in T1).} \label{TableTransactions}
\end{table}

\begin{table}[ht]
\centering
\scalebox{0.8}{
\begin{tabular}{p{3cm}p{5cm}}%{l|}%{|>m{5cm}|>m{15cm}|}
%\rowcolor{orange}
\hline
 \textbf{User}  & \textbf{Address
}\\\hline
U1&A,C,D,G\\
\hline 
U2&B,E\\
\hline
U3&  -\\
\hline 
U4&F\\
\hline
%\hline
\end{tabular}}
\caption{User grouping based on the addresses inferred from Table 
\ref{fig:transacciones}. User 3 (U3) has no assigned addresses, as it corresponds to user 1 (U1).} \label{TableAddress}
\end{table}

Building on this methodological foundation, we conduct a study on the structure of connected components and directed assortativity, as reported by Vallarano \textit{et al.} \cite{Vallarano}, for the first decade of life of the Bitcoin network, and extend the analysis into the second decade of life (see Section \ref{sec:Results2}). In addition, we apply the PageRank and HITS algorithms to offer new insights and a complementary structural perspective on the network (see Sections \ref{sec:pagerank} and \ref{sec:HITS}).

\newpage

\subsection{Limitations of the Reconstruction Heuristics}\label{sub:limitations}

Although there are more advanced heuristics \cite{Zhang2025} that aim to mitigate deliberate traceability-evasion techniques (such as CoinJoin, PayJoin, Peeling , or CoinSwap), our choice of a well-established and limited set of rules allows us to maintain a controlled and consistent analytical scope aligned with the BUN proposed by Vallarano \textit{et al.} \cite{Vallarano}. Therefore, our inferred user clusters should be interpreted as tentative approximations of real-world entities. Future work will address more advanced clustering heuristics.\\

\section{PageRank of the BUN}
\label{sec:pagerank}
PageRank is an algorithm developed by Brin \textit{et al.} \cite{Brin} , originally used to evaluate the importance of Web pages within a hyperlink structure. Its central premise is simple but powerful: A page is important if other important pages link to it. To achieve this, the algorithm models the behavior of a “random surfer” who browses the Web following links and occasionally jumping to a random page. This idea translates into a mathematical problem of Markov chains, for which an eigenvector is calculated that represents the stationary probability of visiting each node in the network\cite{Langville}. Over time, applications expanded beyond the web. Its mathematical foundation, based on graph structures and network theory, allowed it to be adapted to complex networks. \\

\begin{figure}
\centering
\captionsetup{justification=centering}
\includegraphics[scale=0.20]{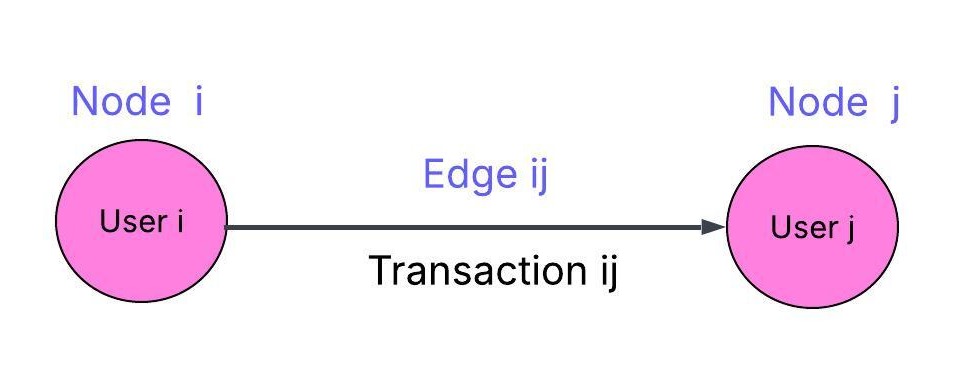}
\caption{Representation of the Bitcoin User Network (BUN) elements, where users $i$ and $j$ are modeled as nodes, and each edge $ij$ denotes a value transfer (transaction $ij$) between them.}
\label{fig1BUNscheme}
\end{figure}

PageRank measures the importance or incoming structural influence of each node within the network\cite{Hajarathaiah}. 
In our network, we define BUN users (user $i$ and user $j$) as nodes (user $i$ and user $j$) and edges (edge $ij$) as transfers (transaction $ij$) between them (see Figure \ref{fig1BUNscheme}). A node with a high PageRank is an influential receiving user, meaning it receives transfers from many addresses or from addresses that are already important. \\

We construct a directed  unweighted graph $G = (V,E)$ that consists of a non‑empty set $V$ of  nodes (or users) and a set $E$ of directed edges (or transactions). Each edge from user $j$ to user $i$ simply denotes the existence of a transaction, without considering its value.
We calculate the centrality metrics (PageRank) \cite{ZhangPR} and the inequality (Gini coefficient)\cite{Vincent}. The data comes from our own BUN database, processed weekly. Each node distributes its influence equally among the nodes it points to (its neighbors). A node that receives many incoming links from important nodes gains a higher PageRank.\\

We define the PageRank of a node \cite{ZhangPR} $i$ as \\
\begin{equation}
PR(i) = \frac{1-d}{N} + d \sum_{j \in M_i} \frac{PR(j)}{L(j)}.
\end{equation} In this equation, $PR(i)$ and $PR(j)$ denote the PageRank values of the nodes $i$ and $j$, respectively, while $d$ is the damping factor, for which we use a default value of 0.85. The term $N$ indicates the total number of nodes in the network. The set $M_i$ consists of all nodes that point to node $i$, while $L(j)$ represents the number of outgoing links from node  $j$.\\

We calculate the Gini coefficient on PageRank values ($G_{c}$) to measure inequality \cite{Vincent} 

\begin{equation}
G_c = \frac{\sum_{i=1}^{n} \sum_{j=1}^{n} \left| x_i - x_j \right|}{2n^2 \bar{x}},
\end{equation} where, $x_i$ represents PageRank of node $i$; $n$ is the total number of nodes; $\bar{x}$ is the average PageRank. The indicator value ranges from 0 to 1. A result close to 0 suggests an even distribution of importance among the nodes, while a value close to 1 indicates a high concentration, in which a few nodes accumulate most of the importance within the network in terms of incoming transactions.\\

Along these lines, analyzing $G_{c}$ would allow a more in-depth characterization of structural inequality from a perspective associated with the “importance” or “influence” of nodes—a particularly relevant dimension in directed networks. Since PageRank evaluates nodes according to the number of transactions links passing through the network,  independent of their value, it is particularly effective in identifying dominant hubs that concentrate the flow of transaction and, therefore, may provide indirect insights into structural concentration patterns that are commonly associated with systemic vulnerability in highly concentrated networks. PageRank is crucial because it can differentiate the importance between nodes with similar degrees or with similar numbers of connections. For example, a node in the IN-Component — the set of nodes from which the strongly connected component (SCC) can be reached but that are not reachable from it — that receives many transactions from many low PageRank peripheral nodes will have a lower total PageRank than a node in the SCC. 
A SCC is defined as a maximal subgraph in which every pair of nodes is mutually reachable through directed paths. By adopting this definition, we can refine our understanding of hierarchy beyond what is captured by simple connectivity.

\section{HITS of the BUN}\label{sec:HITS}

The HITS algorithm (Hyperlink-Induced Topic Search), designed by Jon M. Kleinberg \cite{Kleinberg}, is a set of algorithmic tools originally developed to analyze and extract information from the link structures of hyperlinked environments, such as web pages.
The algorithm is based on two fundamental concepts: hubs and authorities.
\begin{itemize}
       \item Authorities ($x$) are pages considered the best answers for a given topic. A good authority is a page that is referenced by many good hubs.
       \item Hubs ($y$) are pages that act as resource lists, that is, pages that point to multiple relevant authoritative pages. Therefore, a good hub is one that points to many good authorities.
\end{itemize} 

In the HITS algorithm, the analysis of links between pages is essential, as hyperlinks from a page $p$ to a page $q$ confer, to some extent, authority to $q$. As a uniform starting point, we define a vector $z = (1,1,1,\dots,1) \in \mathbb{R}^n$, where $n$ is the number of pages in the focused subgraph ($G_s$). We set the initial authority weight vector ($x_0$) equal to $z$ ($x_0 := z$), and the initial hub vector ($y_0$) equal to $z$ ($y_0 := z$). In other words, the HITS algorithm begins by assuming that each page of $G_s$ has the same initial weight (value 1) as both a potential authority and a potential hub, before the iterative algorithm begins to differentiate their scores through alternating mutual reinforcement operations.\\

The iterative algorithm consists of two operations: updating the authority (Operation I) and updating the hub (Operation II). These operations are applied alternately over $k$ iterations ($k \approx 20$) \cite{Kleinberg}:\\

Operation I: the authority weight $x(p)$ of a page $p$ is updated as the sum of the hub weights $y(q)$ of all pages $q$ that link to $p$

\begin{equation}
x(p) \gets \sum_{q:(q,p)\in E} y(q) .
\end{equation}

Operation II: the hub weight $y(p)$ of a page $p$ is updated as the sum of the authority weights $x(q)$ of all pages $q$ to which $p$ links

\begin{equation}
y(p) \gets \sum_{q:(p,q)\in E} x(q).
\end{equation}

After each update operation, the weight vectors are normalized to ensure that the sum of their squares equals 1. Once the $k$ iterations are completed, the pages with the $c$ largest coordinates in $x_k$ are reported as authorities, and the pages with the $c$ largest coordinates in $y_k$ are reported as hubs (typically $c$ is between 5 and 10).
We apply this HITS algorithm analogously to our BUN. Specifically, we focus on identifying hubs (nodes that point to many authorities and serve as key transmitters) and authorities (users or nodes that receive the majority of important transactions), based on the link structure of our BUN. In this paper, we distinguish between incoming structural influence (captured by PageRank and HITS authority) and outgoing brokerage capacity (captured by HITS hub score).

\section{Volatility and blind experiment}\label{sec:vol}

Volatility is a statistical measure of dispersion based on cryptocurrency prices rather than structural network evolution. Although volatility is not a pure network metric, its inclusion allows us to explore the interaction between structural network dynamics and external market behavior. We calculate volatility based on the volatility estimation of Wang \textit{et al.} \cite{WANG} , but increasing the sampling frequency to obtain more statistics on the fluctuation of the decentralized ledger in the market.\\

\begin{equation}
\mathrm{VOL}_1 = \mathrm{STD}\left( \ln\left(\frac{P_1}{P_0}\right), \ln\left(\frac{P_2}{P_1}\right), \ldots, \ln\left(\frac{P_{1440}}{P_{1439}}\right) \right) \cdot \sqrt{1440},
\end{equation}\\ where, $VOL_1$ is the daily volatility index, STD is the sample standard deviation, ln is the natural logarithm and $\frac{Pi+1}{Pi}$ is the ratio between consecutive spot closing prices at time $i$. We calculate the index as the logarithmic percentage change taken from measurements obtained at the closing price at one-minute intervals. The settlement price is calculated from 1440 samples collected over a 24-hour period. These samples correspond to one closing price recorded per minute for each of the 60 minutes in an hour. Given that a full day consists of 24 hours, this methodology results in a total of 1440 data points or samples used for the settlement calculation. We employ this one-minute price acquisition frequency as it provides greater insight into the variability and dynamics of price movements throughout the day.\\

To have a comprehensive assessment of the cryptocurrency market during external factors or events, it is necessary to understand how volatile Bitcoin is and how it evolves over time. We perform a blind rolling volatility test, where the statistical procedure is conducted independently of any prior labeling of external factors. To do this, we calculate each daily volatility index and apply a statistical test (the Wilcoxon signed-rank Test \cite{Hollander,Gibbons}) to see if the volatility changes one week before a particular day $D$, studying the week following that day in question. Here, $D$ is a variable that we shift day by day. The analysis covers the period from 2020, when high-frequency, high-liquidity BTC spot data for both USDT and EUR became consistently reliable across centralized exchanges, minimizing microstructural distortions, up to November 2025, at which point the study was intentionally concluded.\\

In \ref{appendixA:Volatility}, we detail the acquisition of volatility data and the application of the statistical test. The relevant data and observations can be found in the results (see Section \ref{sec:Results2}).

\newpage

\section{Results}\label{sec:Results2}

In this section, we present the results obtained by processing  
the entire Bitcoin blockchain through a fully synchronized entire node, applying heuristics, and construct our own BUN. We analyze disconnected components, the evolution of the largest connected component (LCC) of the BUN, and the evolution of the ratio between the size of the largest connected component (LCC) and the size of the second largest connected component (second LCC) from 2011 up to late 2025, extending the analysis previously conducted by Vallarano \textit{et al.} \cite{Vallarano} for the period 2011–2018. In addition, we analyze assortativity, examine the evolution of Bitcoin’s price volatility using high-frequency market data, and apply complementary centrality measures based on the PageRank and HITS algorithms. \\

\subsection{Analysis of disconnected components from 2011 to 2025}

\begin{figure*}[ht]
\centering
\captionsetup{justification=centering}
\includegraphics[width=0.75\textwidth]{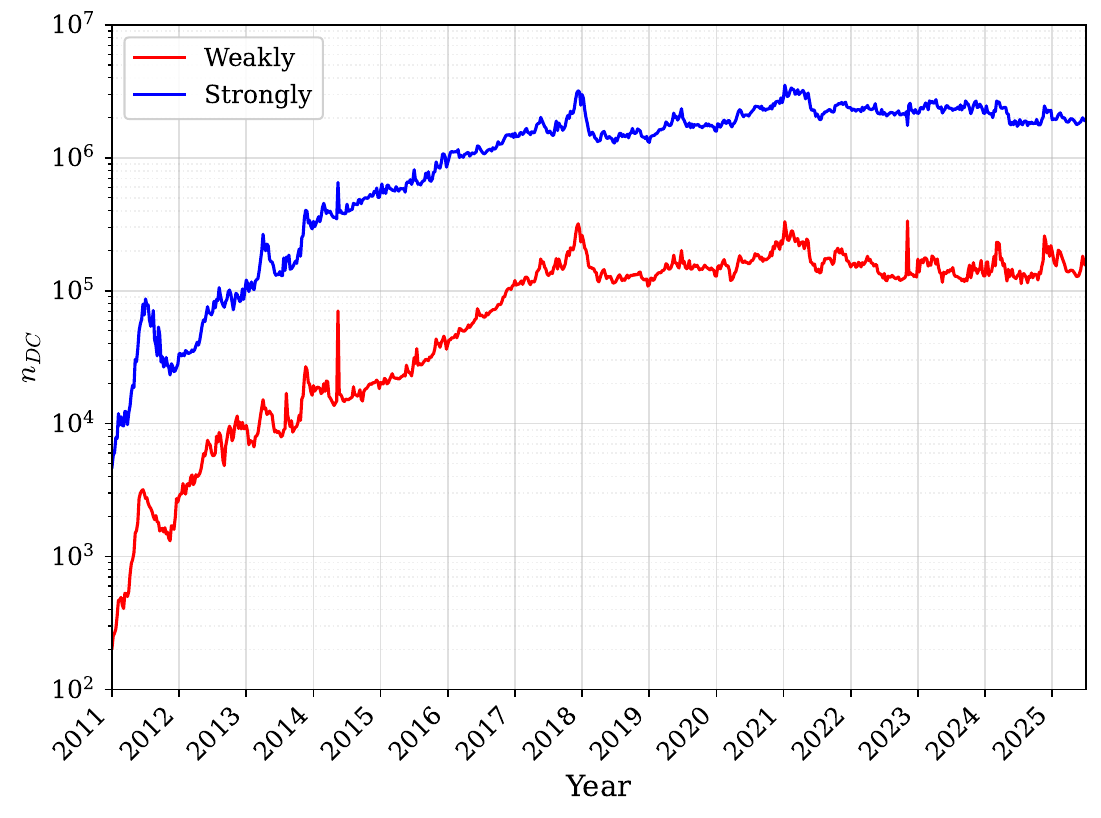}
    \caption{Evolution of the number of weakly (red) and strongly (blue) disconnected components ($n_{DC}$) in the BUN, from 2011 to 2025.}
    \label{fig:EvolutionOfDisconnectedComponents}
\end{figure*}

In Figure \ref{fig:EvolutionOfDisconnectedComponents}, we show the temporal evolution of the number of disconnected components of the BUN ($n_{DC}$) on a logarithmic scale. We calculate the measurements on a weekly basis, covering the period between 2011 and 2025, and distinguish between weakly disconnected components (red line) and strongly disconnected components (blue line). As evidenced in the study by Vallarano \textit{et al.}\cite{Vallarano} (2011–2018), we obtain a similar result in this work. We observe a high degree of topological disconnection in the early days of the Bitcoin network. Specifically, since 2011, the orders of magnitude for the weakly disconnected are $10^2$, while for the strongly disconnected they are $10^3$. Likewise, we find that by 2018, the orders of magnitude for the weakly disconnected approach $10^5$, and those for the strongly disconnected approaches $10^6$. 
At the end of 2018, we observe that the value remains stable, similar to the last values in that year, albeit with minimal fluctuations. This suggests a topological stability in terms of component fragmentation from 2018 onward, although this stability does not extend to all mesoscopic properties, as evidenced by the LSCC dynamics between 2022 and 2024, revealing that global topological stabilization can coexist with local structural turbulence in large-scale networks. However, this stationarity does not extend to all mesoscopic properties of the network. The Bitcoin transaction network is in a phase of topological maturity. The topological stability of the number of disconnected components indicates that the network reaches a state of stationarity in terms of its topological fragmentation pattern. The presence of small fluctuations suggests that the network is not undergoing drastic topological reconfigurations. The topological stability of the number of disconnected components indicates that the dynamics of the network is not driven by intense periods of speculative growth, requiring the massive formation or dissolution of isolated components. \\

\begin{figure*}[!t]
\centering
\captionsetup{justification=centering}
\includegraphics[width=0.75\textwidth]{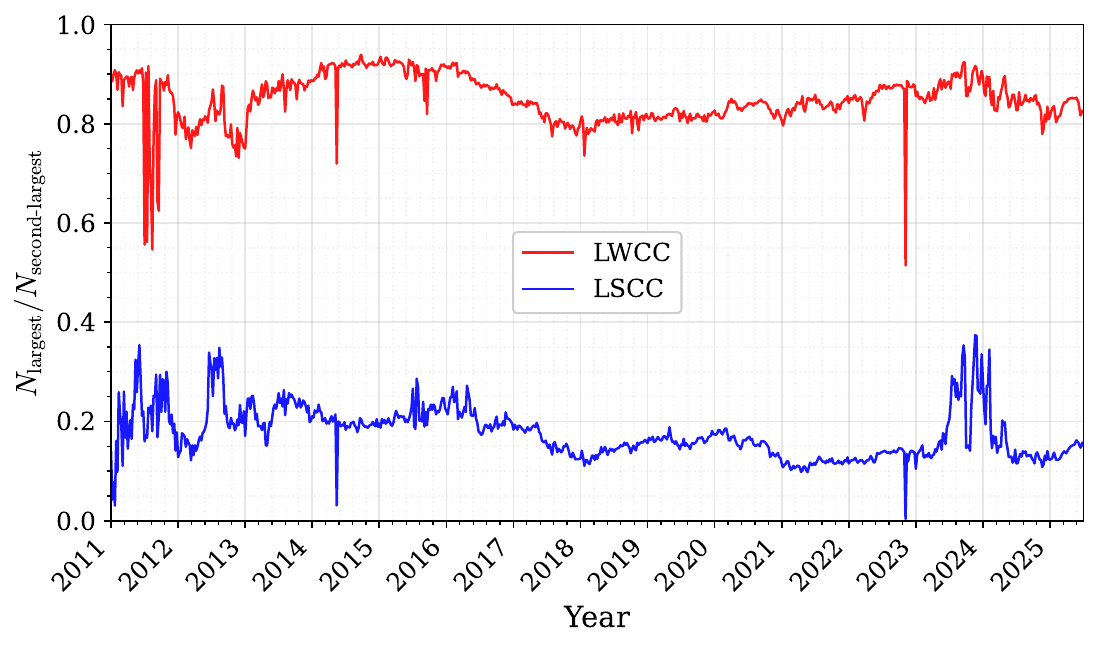}
    \caption{Relative size of the largest weakly connected component (LWCC) and largest strongly connected component (LSCC) from 2011 up to late 2025.}
    \label{fig:EvolutionOfRelativeSize}
\end{figure*}

In Figure \ref{fig:EvolutionOfRelativeSize}, we show the largest weakly connected component (LWCC) and the largest strongly connected component (LSCC). The relative size is calculated by dividing the size of the component (NLCC) by the total number of nodes (N). The relative size of the LWCC remains between 80\% and 90\% of the total number of nodes during 2011 to 2018, similar to Vallarano et al.\cite{Vallarano} which remains stable at around 80\% of the total number of nodes. This result indicates that the vast majority of Bitcoin users are, at least indirectly, connected to each other. Regarding the LSCC, its relative size ranges from 10\% to 30\% of the total number of nodes, partially overlapping with the range reported by Vallarano \textit{et al.}~ \cite{Vallarano} for the first decade. However, during the (2023–2024) period, we observe deviations beyond this historical range, with LSCC values temporarily reach very small relative sizes (without ever vanishing), followed by transient expansions reaching peaks close to 40\%. This does not imply a loss of global connectivity of the weakly connected graph, but rather a phase of mesoscopic turbulence under the adopted temporal aggregation.\\

Finally, there is a return to stabilization between 10\% and 20\% during the (2024-2025) period, indicating that the network has moved from a state of extreme turbulence to one of dynamical instability.

\subsection{Evolution of the ratio between the size of the largest connected component (LCC) and the size of the second largest connected component (second LCC)}

\begin{figure*}[!ht]
\centering
\captionsetup{justification=centering}
\includegraphics[width=0.75\textwidth]{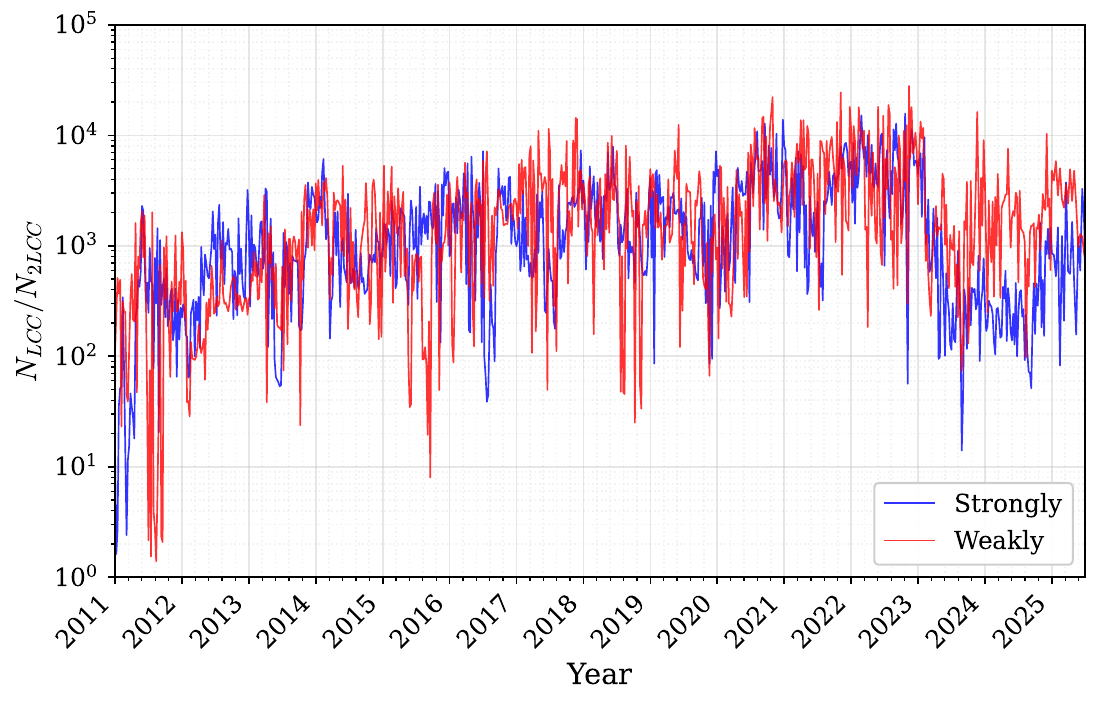}
    \caption{Evolution of the ratio between the size of the largest and second-largest components (weak and strong) from 2011 up to late 2025.}
    \label{fig:EvolutionOfRatio1stLCC&2ndLCC}
\end{figure*}

In Figure \ref{fig:EvolutionOfRatio1stLCC&2ndLCC}, we show the evolution of the ratio between the LWCC (or LSCC) and the second largest component. From 2011–2018, in agreement with Vallarano et al.\cite{Vallarano}, the largest component is about three orders of magnitude larger and clearly dominates the network. From 2018 to late 2025, this ratio increases further—by three to four orders of magnitude in the weak case and two to three in the strong one—reinforcing the persistence of a few large-scale components that remain mesoscopically stable over time.

\subsection{Analysis of assortativity}
Assortativity is a measure that describes the mixing patterns of the network. Assortativity occurs when nodes with high (low) degrees tend to connect with other nodes with high (low) degrees. In contrast, disassortativity occurs when nodes with high degrees (hubs) tend to connect with nodes with low degrees, and vice versa. From the BUN, we calculate four variants of the assortativity coefficient ($r$), correlating the in-degree and out-degree:

\begin{itemize}
       \item $r(out,out)$: correlation between out-degree and out-degree. 
       \item $r(out, in)$: correlation between out-degree and in-degree.   
       \item $r(in, out)$: correlation between in-degree and out-degree. 
       \item $r(in,in)$: correlation between in-degree and in-degree.
\end{itemize}

\begin{figure*}[!ht]
\centering
\captionsetup{justification=centering}
\includegraphics[width=0.75\textwidth]{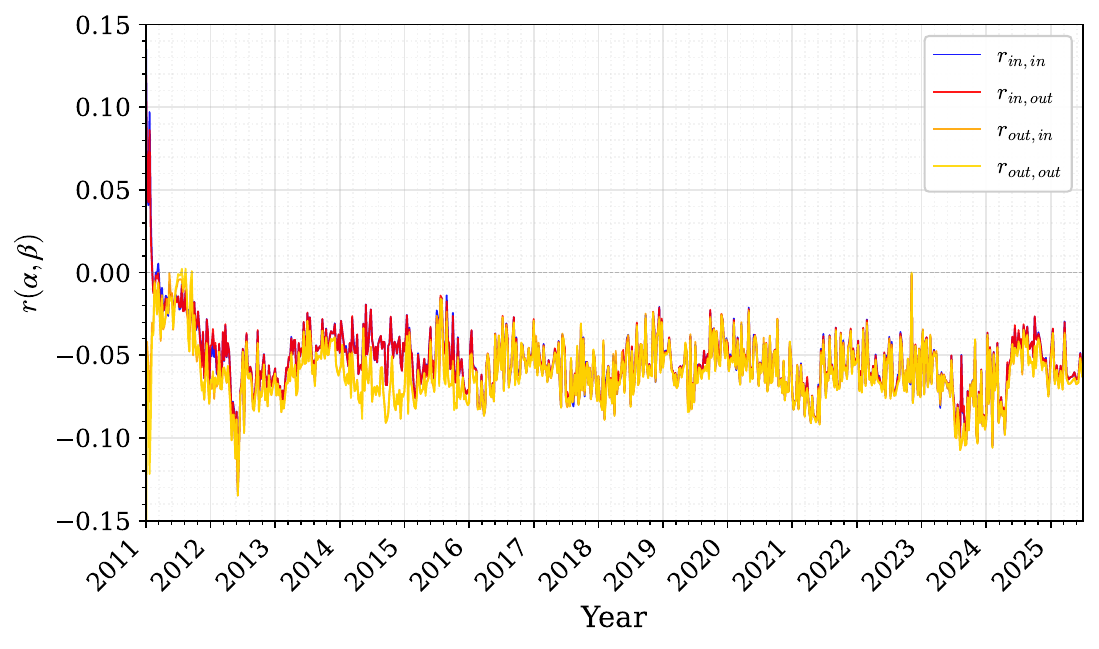}
    \caption{Evolution of Newman's four directed assortativity coefficients ($r_{\text{out,out}}$, $r_{\text{out,in}}$, $r_{\text{in,out}}$, $r_{\text{in,in}}$) for the BUN from 2011 up to late 2025.}
    \label{fig6:Coeficientes de Newman}
\end{figure*}

In Figure \ref{fig6:Coeficientes de Newman}, we show the evolution of the four Newman's assortativity coefficients \cite{Newman} of the BUN over time, from 2011 up to November 2025. In this figure, we examine how the connectivity of nodes (users) correlates with their in-degrees and out-degrees, revealing the weakly disassortative nature of the BUN (its values are negative and remain so throughout the period). This disassortative characteristic suggests that users (nodes) with a large number of transactions (high degrees) tend to connect with users who have a small number of transactions (low degrees). These results are similar to those obtained by Vallarano et al. \cite{Vallarano}, reinforcing the evidence that this disassortative characteristic has remained remarkably topologically stable throughout the second decade of life of Bitcoin.\\

\newpage
\subsection{Analysis of PageRank of the BUN}

\begin{figure}[!ht]
\centering
\captionsetup{justification=centering}
\includegraphics[width=0.75\textwidth]{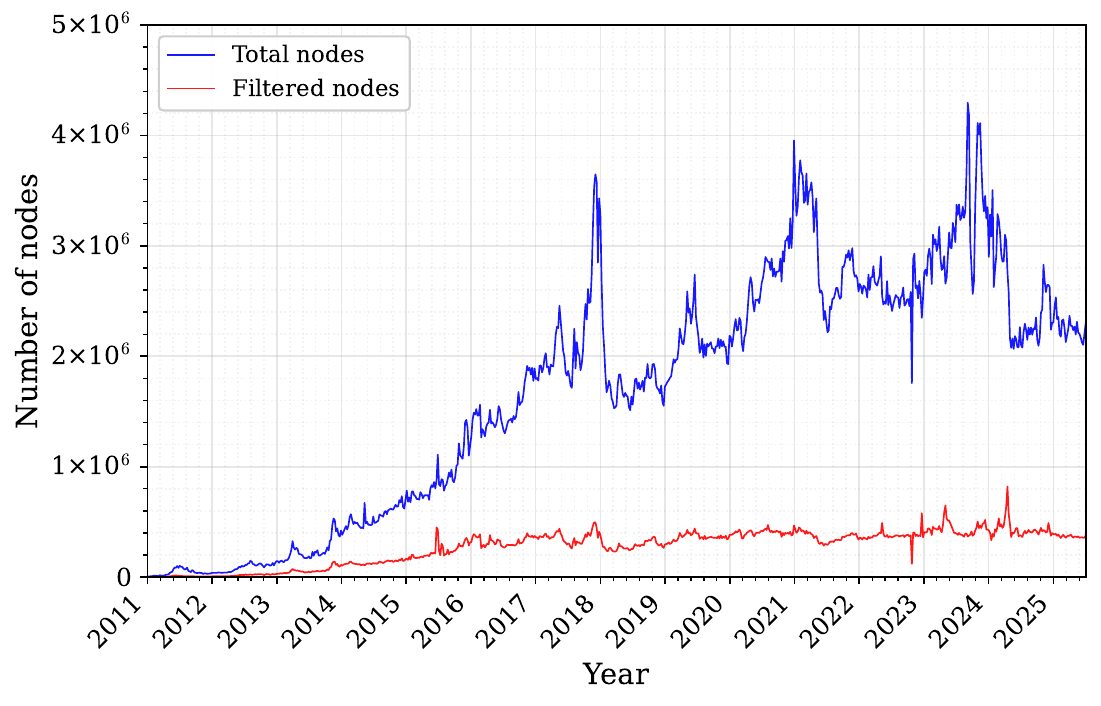}
    \caption{Growth of total nodes in the BUN versus nodes with degree greater than 2 (filtered nodes) from 2011 up to late 2025.}
    \label{fig:StructuralGrowthBTC}
\end{figure}

Figure \ref{fig:StructuralGrowthBTC}  shows the structural growth of Bitcoin, displaying the total number of nodes (in blue) and the number of filtered nodes (in red), which we define in this analysis as nodes with a degree greater than 2. The total number of nodes increases steadily, while the subgraph of active nodes with a degree greater than 2 exhibits slower and more stable growth. More than 60\% of the nodes have a degree of 2 or less, indicating a network with many peripheral participants and few central hubs. 

\begin{figure*}[!t]
\centering
\captionsetup{justification=centering}
\includegraphics[width=0.75\textwidth]{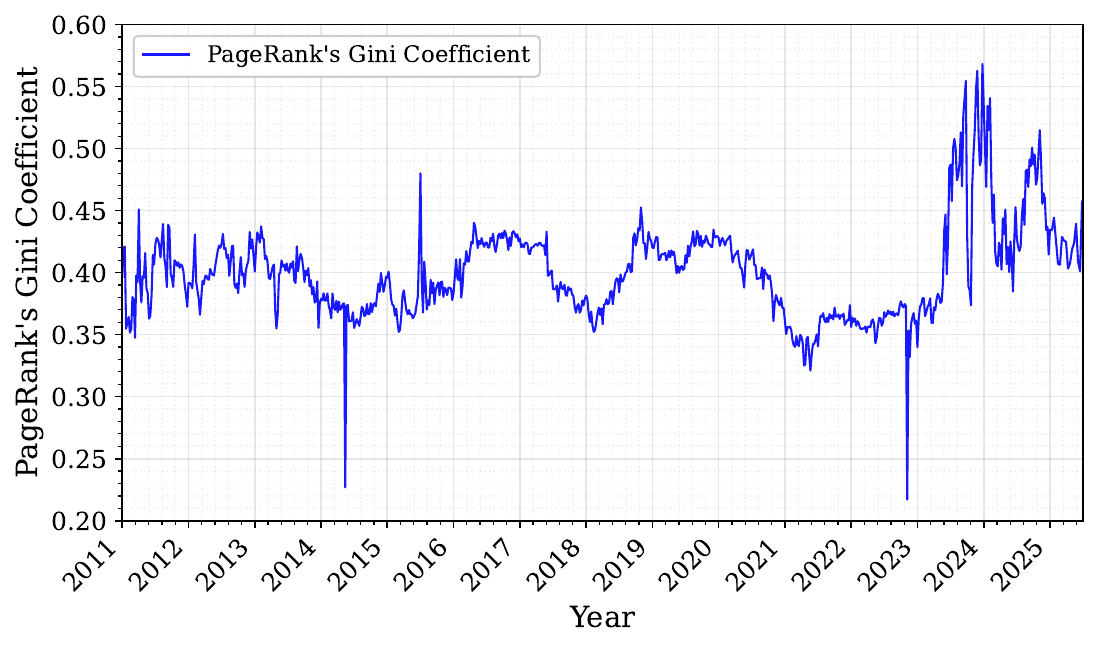}
    \caption{Gini coefficient of PageRank centrality values from 2011 up to late 2025.}
    \label{fig:PageRank&Gini}
\end{figure*}

The analysis of the BUN up to late 2025, incorporating metrics such as PageRank and the Gini Coefficient (Figure \ref{fig:PageRank&Gini}), reinforces the mesoscale conclusions of Vallarano \textit{et al.}\cite{Vallarano}(2011–2018) regarding structural concentration. The concentration of influence (as measured by PageRank) and the inequality in its distribution (reflected in consistently high Gini values, which peak at 0.55 in 2024 and then fluctuate between that peak and 0.4 in the following years) reveal the existence of a few large-scale connected components that function as the critical backbone of the system.

\subsection{Analysis of HITS of the BUN}

\begin{figure*}[!t]
\centering
\captionsetup{justification=centering}
\includegraphics[width=0.75\textwidth]{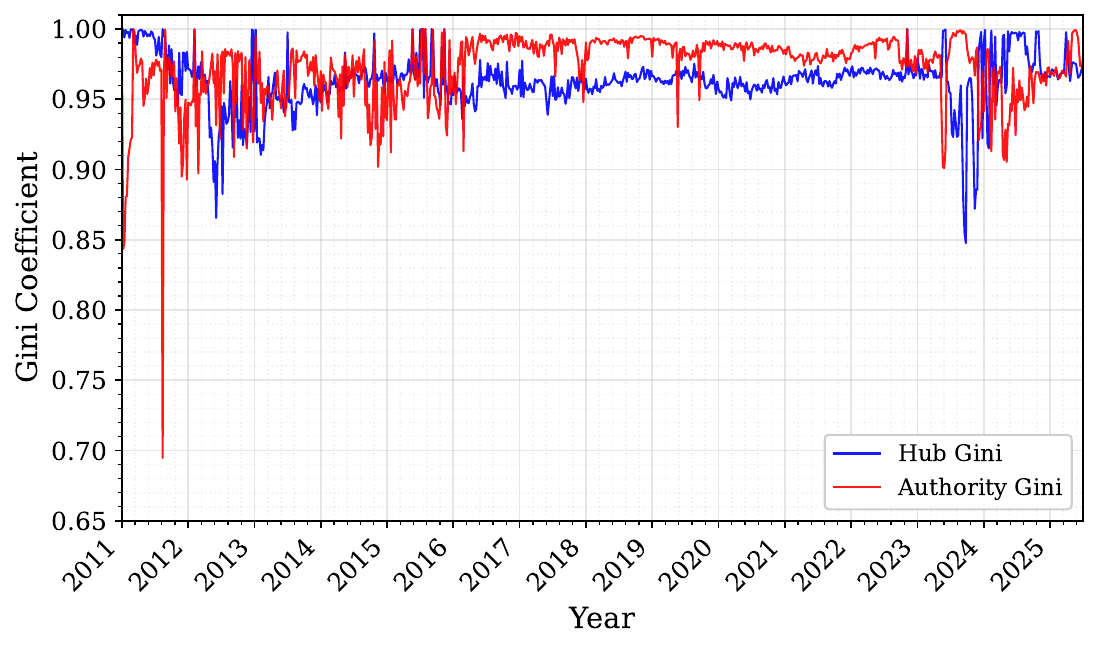}
    \caption{Gini coefficients of HITS hub and authority scores for the BUN from 2011 to 2025, showing extreme inequality in both metrics.}
    \label{fig:Hits}
\end{figure*}

Applying the HITS algorithm yields a Gini coefficient that exceeds 0.9 (see Figure \ref{fig:Hits}) , indicating an extreme concentration of structural influence, both in terms of authority (incoming) and hub (outgoing) roles. This results in a highly concentrated multi-hub topology with extreme core–periphery characteristics, which in certain periods approaches a star-like configuration due to the overwhelming dominance of a very small subset of nodes
, while still remaining embedded in a globally multi-hub structure. Practically, despite millions of users, a very large fraction of transaction links is concentrated around a few dominant hubs, which are often associated in the literature with large exchanges and custodial services, analogous to urban systems in which a very large proportion of transactions concentrate around a limited number of dominant commercial hubs. Although Vallarano \textit{et al.} \cite{Vallarano} emphasize multiple powerful hubs that form locally star-shaped substructures, our approach complements this by highlighting the directionality of influence and the pronounced asymmetry in transactional flows.\\

\subsection{Evolution of the daily volatility index from 2020 to November 2025}\label{sec:hyper}
\begin{figure*}[!ht]
\centering
\captionsetup{justification=centering}
\includegraphics[width=0.75\textwidth]{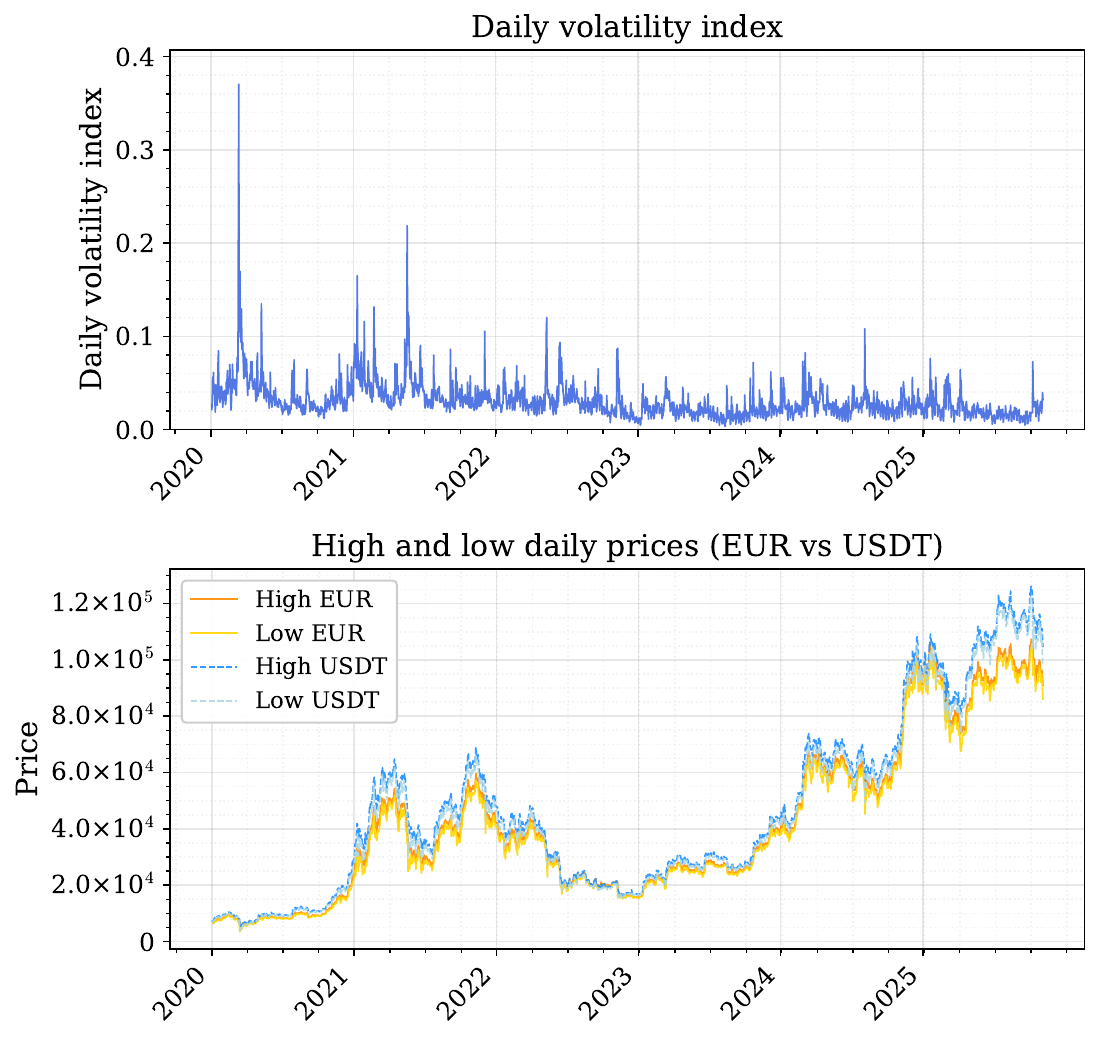}
\caption{Daily Volatility index of Bitcoin (top) and daily high-low price ranges in EUR and USDT (bottom), from 2020 to November 2025.}
\label{fig:dailyvolatility}
\end{figure*}

We calculate the daily volatility index using one-minute closing prices (1440 samples per day). For visualization purposes, we additionally plot daily high–low price ranges in USDT and EUR to provide an intuitive comparison with the intraday volatility estimator. Figure \ref{fig:dailyvolatility} shows two graphs: one illustrating the daily volatility index, and the other comparing the daily high and low prices in both EUR and USDT during this period. We can observe a trend of the maximum amplitude peaks of the volatility indices. As years go by, the maximum peaks of the daily volatility indices tend to decrease and deviate less from their mean value, maintaining their standard deviation during recent years (see Table \ref{AppendixTable3C}).\\

Hypervolatility (or excess volatility) is characterized by an exceptionally high level of price fluctuations and a pronounced instability of the volatility itself. According to Baur \textit{et al.} \cite{Baur}, Bitcoin exhibits excess volatility, with price fluctuations approximately ten times higher than those of the main fiat currencies, such as the US dollar against the euro and the yen. However, Bitcoin continues to remain relatively hypervolatile in the cryptocurrency market, and there is a clear trend toward lower volatility intensity as time goes by. This behavior can also be partly attributed to fluctuations in the fiat currency markets.\\

We apply the Wilcoxon signed-rank Test \cite{Hollander,Gibbons} to evaluate the days on which an event $D$ is assumed to occur. To avoid bias, we perform a sweep of the event day across the entire study period and analyze the volatility values on the days preceding and following each assumed event. Upon applying the statistical test, we do not detect statistically significant changes in the daily volatility index that can be directly attributed to the events $D$. This does not exclude the existence of such effects under alternative metrics. An interesting direction for future work would be to explore how the mesoscopic properties of the BUN behave in proximity to specific socio-political and economic events.\\

\newpage

\FloatBarrier
\section{Conclusions}

This study provides a long-term structural analysis of the Bitcoin User Network from 2011 up to late 2025 based on a complete full-node dataset and a reproducible address-clustering methodology. The extended temporal window and the incorporation of direction-sensitive centrality measures allow us to identify clear mesoscopic trends that characterize the evolution of the network beyond its first decade.
The results show that Bitcoin’s transactional activity becomes progressively concentrated within a few large-scale structures. The number of disconnected components stabilizes after 2018 (Figure~\ref{fig:EvolutionOfDisconnectedComponents}), while the percentage of users belonging to the largest weakly connected component remains persistently high, between roughly 80\% and 90\% (Figure~\ref{fig:EvolutionOfRelativeSize}). The ratio between the largest and second-largest components increases by several orders of magnitude in the second decade (Figure~\ref{fig:EvolutionOfRatio1stLCC&2ndLCC}), confirming the emergence of a dominant backbone.\\

The connectivity patterns within this dominant component remain consistently hierarchical. All directed assortativity coefficients stay negative throughout the entire period (from 2011 to late 2025) (Figure \ref{fig6:Coeficientes de Newman}), revealing persistent disassortative mixing in which high-degree nodes interact primarily with low-degree users. Beyond the mesoscopic descriptors known in the literature \cite{Vallarano}, our application of direction-sensitive centrality algorithms—PageRank and HITS—constitutes a key contribution of this work, as they allow us to perform centrality metrics, considering both incoming and outgoing transaction flows, and providing a complementary and more qualitative characterization of structural influence and transactional roles within the system. The distribution of structural influence reflects this hierarchy: the proportion of structurally active nodes grows much more slowly than the total number of users (Figure \ref{fig:StructuralGrowthBTC}), and the centrality of PageRank exhibits long-term inequality, with Gini coefficients between 0.4 and 0.55 (Figure \ref{fig:PageRank&Gini}). Directional asymmetries become even more pronounced under the HITS algorithm, where Gini coefficients greater than 0.9 for authority and hub scores (Figure~\ref{fig:Hits}) show that an extremely small subset of nodes concentrates both incoming and outgoing structural roles.\\

Last but not least, these observations reveal a clear and strengthening pattern of concentration within distribution: although Bitcoin’s protocol is decentralized, the user network evolves toward a mesoscopic configuration dominated by a few large-scale
connected components that function as the critical backbone of the system. This structure produces a persistent and asymmetric core–periphery topology, characterized by a high influence inequality and directional concentration. These findings highlight the importance of studying Bitcoin not only from the perspective of its decentralized architecture but also through the emergent organizational patterns that arise from user behavior. Future work may refine the reconstruction of users using more advanced heuristics and investigate how these concentrated structures relate to systemic risk in proximity to specific socio-political and economic events.

\section*{Credit authorship contribution statement}
Myriam Nonaka: Methodology, Software, Formal analysis, Investigation, Data curation, Visualization, Writing – Original Draft, Writing – Review \& Editing; F. Javier Marín-Rodríguez: Software, Investigation, Visualization, Writing – Original Draft, Writing – Review \& Editing; Alexander Jiricny: Software, Data curation, Investigation, Visualization, Writing – Original Draft; Miguel Romance: Conceptualization, Methodology, Project administration, Funding Acquisition, Supervision; Regino Criado: Conceptualization, Methodology, Project administration, Funding Acquisition, Supervision; Sergio Iglesias-Pérez: Methodology, Resources, Validation; Alberto Partida: Conceptualization, Supervision, Project administration, Methodology, Writing – Review \& Editing. Alberto Partida led the overall direction of the project and provided strategic and scientific guidance throughout the research.

All authors contributed to discussions during the preparation of the manuscript and approved the final version.

\section*{Funding}
This work has been supported by INCIBE/URJC Agreement M3386/2024/0031/001 within the framework of the Recovery, Transformation and Resilience Plan funds of the European Union (Next Generation EU).

\section*{Declaration of competing interest}
The authors declare that they have no known competing financial interests or personal relationships that could have appeared to influence the work reported in this paper.

\section*{Declaration of Generative AI and AI-assisted Technologies in the Writing Process}
During the preparation of this manuscript, the authors used ChatGPT, DeepL Translator, and the Overleaf AI Assistant to support language editing, improve clarity and phrasing, and enhance overall readability. After using these tools, the authors carefully reviewed and revised the content, and they take full responsibility for the final version of the manuscript and its scientific integrity.
\appendix
\section{Statistical test and volatility data acquisition}
\label{app1}

\label{appendixA:Volatility}
\subsection{Volatility data acquisition}
We use the Binance API \cite{BinanceAPI} to obtain weekly Bitcoin price data in USDT and EUR over time. We construct Table \ref{AppendixTable3C} using the daily volatility data presented in Figure \ref{fig:dailyvolatility}, reporting the mean, maximum, minimum, and standard deviation for the period from 2020 to November 2025.

\begin{table}[ht]
\centering
\scalebox{0.8}{
\begin{tabular}{lcccc}
\hline
\textbf{Year} & \textbf{Max} & \textbf{Mean} & \textbf{Std} & \textbf{Min} \\
\hline
2020 & 0.307 & 0.033 & 0.026 & 0.008 \\
2021 & 0.214 & 0.045 & 0.023 & 0.020 \\
2022 & 0.108 & 0.031 & 0.016 & 0.006 \\
2023 & 0.068 & 0.020 & 0.010 & 0.003 \\
2024 & 0.110 & 0.026 & 0.012 & 0.007 \\
2025 & 0.077 & 0.021 & 0.011 & 0.005 \\
\hline
\end{tabular}}
\caption{Daily volatility index from 2020 to November 2025, showing the calculated maximum (Max), minimum (Min), mean (Mean), and standard deviation (Std) values.}
\label{AppendixTable3C}
\end{table}

 \subsection{Statistical test}
The Wilcoxon signed-rank Test \cite{Hollander,Gibbons} is a nonparametric statistical test used to compare two related samples. This kind of statistical test is commonly applied in scenarios such as before-and-after studies to assess the effect of an intervention, which in our case is the effect of an event $D$ on the Bitcoin network. We take as a dataset two groups: one that belongs to the daily volatility values during the week prior to the event ($x_{b}$ ), it means the group of data of daily volatility indices ordered from one day before the event, two days before the event, three days before the event, and so on, and the other that belongs to the daily volatility values during the week following the event ($x_{a}$ ), it means the group of data of daily volatility indices ordered from one day after the event, two days after the event, three days after the event, and so on. The null hypothesis (h) to be evaluated will be that the data ($x_{b}$ -$x_{a}$ ) come from a distribution with a median equal to 0, and the alternative hypothesis is that they come from a distribution with a median different from 0. We will use a significant level of 5\%. That is, if we accept the null hypothesis (h=0), there is no significant difference between both groups after the event occurred. Otherwise, the null hypothesis is rejected (h=1).  \\

\begin{figure}[ht]
\centering
\captionsetup{justification=centering}
\includegraphics[width=0.9 \textwidth]{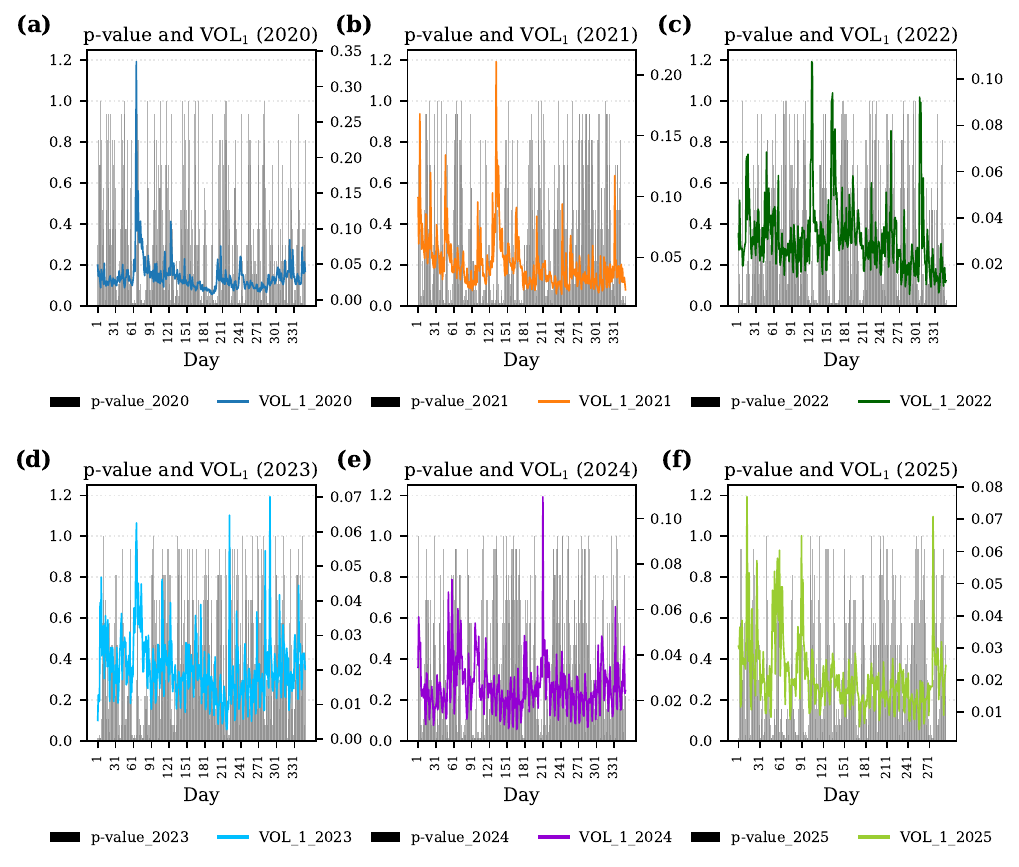}
    \caption{$VOL_1$ of BTC from 2020 to November 2025: (a), in color blue, the $VOL_1$ graph of year 2020; (b), in color orange, the $VOL_1$ graph  of year 2021;(c), in color green, the $VOL_1$ graph of year 2022; (d), in light blue, the $VOL_1$ graph of year 2023; (e), in color violet, the $VOL_1$ graph of year 2024; (f), in light green, the $VOL_1$ graph  of year 2025. In color gray the p-value associated with the statistical test}   \label{graph1f}
\end{figure}

In Figure \ref{graph1f} of (a) - (f), we show in color the graphs of daily volatility indexes $VOL_1$ from 2020 to November 2025. At the same time, we plotted the p-value associated with the statistical test in gray.

%%       \citet{<label>} ==> Lamport (1994)
%%       \citep{<label>} ==> (Lamport, 1994)
%%




%%
%% End of file `elsarticle-template-harv.tex'.
\end{document}